\documentstyle[aps,prl,multicol]{revtex}
\renewcommand{\baselinestretch}{1}
\begin{document}
\title{Minimum cbits for remote preperation and measurement of a qubit}
\author{Arun K. Pati$^{(1)}$}
\address{School of Informatics, University of Wales, Bangor LL 57 1UT, UK}
\address{$^{(1)}$Theoretical Physics Division, 5th Floor, C.C., BARC, Mumbai-400085, 
India}

\date{\today}
\maketitle
\def\ra{\rangle}
\def\la{\langle}
\def\ver{\arrowvert}
\begin{abstract}
We show that a qubit chosen from equatorial or polar
great circles on a Bloch spehere can be remotely prepared with 
one cbit from Alice to Bob if they share one ebit of entanglement.
Also we show that any single particle measurement on an arbitrary qubit can be
remotely simulated with one ebit of shared entanglement and
communication of one cbit.
\end{abstract}

\vskip .5cm

PACS           NO:    03.67.-a, 03.65.Bz\\

email:akpati@sees.bangor.ac.uk\\

\vskip 1cm

\begin{multicols}{2}

\par

   The state of a quantum system contains a large amount of information which
cannot be accessed by an observer.  How well one can extract
 and utilise the largely inaccessible 
quantum information is the subject of quantum information theory. One of the
surprising discoveries in this area is the {\em teleportation}  of an 
{\em unknown} quantum state by Bennett {\it et al} \cite{cb} from one place
 to another without ever
physically sending the particle. A qubit, for example, can be sent
from Alice to Bob provided they share an
Einstein-Podolsky-Rosen (EPR) pair and Alice
carries out a Bell-state measurement on the qubit and one half of the EPR
pair, and
sends two bits of classical information to Bob, who in turn can
 perform a unitary operation on his particle to get the original state.
 The quantum teleportation of
 photon has been demonstrated experimentally by Bouwmeester {\it et al} 
\cite{db} and Boschi {\it et al} \cite{dbe}. The continuous version of quantum
teleportation has been also verified by Furusawa {\it et al} \cite{af}.
Though, a qubit contains a doubly
infinity of bits of information, only two classical bits (cbits) are necessary to
transmit a qubit in the teleportation process. This raises the question,
whether it is
really the minimum number of cbits needed to transmit a qubit.
 What about the rest of the infinity of this number of bits? 
It has been suggested that the remaining bits flow across the
entanglement channel \cite{rj}.
Is it that two cbits are required just to preserve 
the causality (the peaceful co-existence of quantum theory and relativity) 
or is it the {\em ``soul''} of an {\em unknown} qubit
 (without which the qubit cannot be reconstructed, the particle is just
 being in a random mixture at Bob's place)?

  Recently several philosophical implications of quantum teleportation and its
experimental verification have been brought out by Vaidman \cite{lv}.  
 Though  quantum teleportation requires a quantum channel which is an 
entangled pair, doubts have been raised whether teleportation is really a
non-local phenomena \cite{sp}.  Hardy \cite{lh} has  argued that one can
 construct a local theory where
cloning of a state is not possible but teleportation is. Interestingly, 
the old issue of mimicking quantum theory by a local hidden variable (LHV)
theory has
been revived by Brassard et al \cite{betal} and Steiner \cite{ms} who
show that non-local correlations of quantum theory can be simulated by
local hidden variable theory with classical communication. A natural question
then is, if classical communication can help in mimicking non-local correlation,
can one teleport a quantum state with extra number of cbits. This has been
answered by Cerf et al \cite{cgm} who have proved that one can construct 
a classical teleportation scheme of a {\em known} state from Alice to Bob with the help
of 2.19 cbits (on an average) provided they have initially shared local hidden variables. This
is an interesting result. They compare the cbits required in classical 
teleportation to cbits required in quantum teleportation and argue that only
.19 bit more is required when one uses local hidden variables.

In this note we show that there is a simple scheme for remote preparation
and measurement of a qubit {\em known} to  Alice but {\em unknown} to Bob.
This requires only one cbit to be transmitted from
Alice to Bob. This may also be called {\em teleportation of a known qubit}.
Unlike the teleportation of an {\em unknown} qubit, here, we do not require
a Bell-state measurement. Only a single particle von Neumann measurement is
necessary. The qubit which is
intended to be transmitted does not play any direct role in the measurement
process except for the fact that it's state is known to Alice.
A qubit chosen from equatorial or polar
great circles on a Bloch spehere can be remotely prepared with 
one cbit from Alice to Bob if they share one ebit of entanglement.
Further we show that any single particle measurement on an arbitrary qubit can be
remotely simulated with one ebit and communication of one cbit.
This also shows that the classical teleportation envisaged by Cerf et al
\cite{cgm} actually requires 1.19 bits more than that of a situation where
one uses entangled pairs rather than local hidden variables. Since they think
of transmitting a {\em known} qubit and in teleportation one sends an
{\em unknown} qubit one should not compare the classical information cost
in the above situation.

Let us consider  a pure input state $\ver\Psi \ra \in {\cal H} = C^2$, 
which is the state of a qubit.
 An arbitrary qubit can be represented as

\begin{equation}
\ver \Psi \ra = \alpha \ver 0 \ra + \beta \ver 1 \ra,
\end{equation}
where we
can choose  $\alpha$ to be real and $\beta$ to be a complex number,
in general (up to $U(1)$ equivalance classes of states).
This qubit can be represented by a point on a sphere $S^2$ (which is the
projective Hilbert space ${\cal P} = CP(1)$ for any two-state system) with the help of two real
parameters $\theta$ and $\phi$, where $\alpha = \cos {\theta \over 2}$ and
$\beta = \sin {\theta \over 2} \exp (i \phi)$. Now Alice wants to transmit the
above qubit to Bob. She can either physically send the particle (which is not
interesting) or she needs to send a doubly infinity of bits of information
 across a
classical channel to Bob. However, as we will show,
 there is a very simple procedure to send the information content of a
 qubit without ever sending it or
without ever sending an infinity of bits of information. Just  one
cbit is required to send the information content of a qubit provided
Alice and Bob share one half of the particles from an EPR source.
The EPR state of 
the particles 1 and 2 is given by,

\begin{equation}
\ver \Psi^- \ra_{12} = {1 \over \sqrt 2}(\ver 0 \ra_1 \ver 1 \ra_2 -
\ver 1 \ra_1 \ver 0 \ra_2).
\end{equation}

Suppose Alice is in possession of 1 and Bob is in possession of 2. The
qubit $\ver \Psi \ra$ is {\em known} to  Alice and  {\em unknown} to 
Bob. Since Alice knows the state she can chose to measure the particle $1$ 
in any  basis she wants. 
Alice carries out measurement on particle
1 by projecting onto the ``qubit basis''
$\{ \ver \Psi \ra, \ver \Psi_{\perp} \ra \}$, where the ``qubit  basis''
 is related to the old basis $\{\ver 0 \ra, \ver 1 \ra \}$
in the following manner

\begin{eqnarray}
\ver 0 \ra_1 = \alpha \ver \Psi \ra_1 - \beta \ver  \Psi_{\perp}  \ra_1
 \nonumber\\
\ver 1 \ra_1 = \beta^* \ver \Psi \ra_1 + \alpha \ver \Psi_{\perp}  \ra_1.
\end{eqnarray}
By this change of basis the normalisation and orthogonality relation
between basis vectors are preserved.
Now writing the entangled state $\ver\Psi^- \ra_{12}$ in
the ``qubit basis'' $\{ \ver \Psi \ra_1 , \ver \Psi_{\perp} \ra_1 \}$ gives us 

\begin{eqnarray}
\ver \Psi^- \ra_{12} = {1 \over \sqrt 2}[ \ver \Psi \ra_1 \ver \Psi_{\perp} \ra_2
 - \ver \Psi_{\perp} \ra_1 \ver \Psi \ra_2 ],
\end{eqnarray}
which is also a consequence of invariance of $\ver \Psi^- \ra_{12}$ under
$U_1 \otimes U_2$ operation. The total state after a 
a single particle von-Neumann measurement 
(if the outcome of Alice is $\ver \Psi_{\perp} \ra_1$) is given by
 
\begin{equation}
\ver \Psi_{\perp} \ra_1 \la  \Psi_{\perp}  \ver \Psi^- \ra_{12}
 =  -{1 \over  \sqrt 2}
 \ver \Psi_{\perp} \ra_1 \otimes \ver \Psi \ra_{2} 
\end{equation}

When
she sends her measurement result
(one bit of classical information)
to Bob, then particle 2 can been found in the
original state   $(\alpha \ver 0 \ra_2 + \beta \ver 1 \ra_2)$
which is nothing but the remote prepartion of a
{\em known} qubit.
If the outcome of Alices's measurement result is $\ver \Psi \ra_1$ then the classical
communication from Alice would tell Bob that he has obtained a state
which is $ \ver \Psi_{\perp} \ra = (\alpha \ver 1 \ra_2 - \beta^*
\ver 0 \ra_2) $. This is a
complement qubit which is orthogonal to the original one. The
resulting  state (if the outcome is
$\ver \Psi \ra_1$) is given by
 
\begin{equation}
\ver \Psi \ra_1  \la  \Psi  \ver \Psi^- \ra_{12}
 =  -{1 \over  \sqrt 2}
 \ver \Psi \ra_1 \otimes \ver \Psi_{\perp} \ra_{2}
\end{equation}

There is nothing special about sharing an EPR singlet state. In fact,
 Alice
and Bob can share any other maximally entangled state  from
the basis $\{\ver \Psi^+ \ra_{12}, \ver \Phi^{\pm} \ra_{12} \}$.
These  can be expressed in terms of the qubit basis as 

\begin{eqnarray}
&& \ver \Psi^+ \ra_{12} = - {1 \over \sqrt 2}[ \ver \Psi \ra_1 (\sigma_z) \ver \Psi_{\perp} \ra_{2} +
\ver \Psi_{\perp} \ra_1 (\sigma_z) \ver \Psi \ra_{2} ]    \nonumber \\
&& \ver \Phi^+ \ra_{12} =  {1 \over \sqrt 2}[ \ver \Psi \ra_1 (i\sigma_y) \ver \Psi_{\perp} \ra_{2} +
\ver \Psi_{\perp} \ra_1 (i\sigma_y) \ver \Psi \ra_{2} ]    \nonumber \\
&& \ver \Phi^- \ra_{12} =  {1 \over \sqrt 2}[ \ver \Psi \ra_1 (\sigma_x) \ver \Psi_{\perp} \ra_{2} +
\ver \Psi_{\perp} \ra_1 (\sigma_x) \ver \Psi \ra_{2} ]   
\end{eqnarray}
where $\sigma_x, \sigma_y$ and $\sigma_z$ are the Pauli matrices.
When Alice and Bob share $\ver \Psi^+ \ra_{12}$
then the resulting states
after  a single particle von Neumann measurement
and classical communication are given by

\begin{eqnarray}
&& \ver \Psi \ra_1  \la  \Psi  \ver \Psi^+ \ra_{12}
=  {1 \over  \sqrt 2}
 \ver \Psi \ra_1 \otimes (\sigma_z)\ver \Psi_{\perp} \ra_{2}  \nonumber \\
&& \ver \Psi_{\perp} \ra_1  \la  \Psi_{\perp}  \ver \Psi^+ \ra_{12} 
 = {1 \over \sqrt 2}
 \ver \Psi_{\perp} \ra_1 \otimes (\sigma_z)\ver \Psi \ra_{2} 
\end{eqnarray}
Similarly, 
when they share   $\ver \Phi^+ \ra_{12}$, then the resulting states
after  a single particle von Neumann measurement
and classical communication are given by

\begin{eqnarray}
&& \ver \Psi \ra_1  \la  \Psi  \ver \Phi^+ \ra_{12}
 = - {1 \over  \sqrt 2}
 \ver \Psi \ra_1 \otimes (i\sigma_y)\ver \Psi_{\perp} \ra_{2} \nonumber \\
&& \ver \Psi_{\perp} \ra_1  \la  \Psi_{\perp}  \ver \Phi^+ \ra_{12}
 = -{1 \over \sqrt 2}
 \ver \Psi_{\perp} \ra_1 \otimes (i\sigma_y)\ver \Psi \ra_{2}
\end{eqnarray}
Finally, 
when they share  $\ver \Phi^- \ra_{12}$, then the resulting states after 
 a single particle von Neumann measurement and
classical communication are given by

\begin{eqnarray}
&& \ver \Psi \ra_1  \la  \Psi  \ver \Phi^- \ra_{12}
 =  {1 \over \sqrt 2}
 \ver \Psi \ra_1 \otimes (\sigma_x)\ver \Psi_{\perp} \ra_{2}  \nonumber \\
&& \ver \Psi_{\perp} \ra_1  \la  \Psi_{\perp}  \ver \Phi^- \ra_{12}
 = {1 \over  \sqrt 2}
 \ver \Psi_{\perp} \ra_1 \otimes (\sigma_x)\ver \Psi \ra_{2}
\end{eqnarray}

 In general if Alice finds $\ver \Psi_{\perp} \ra_1$
in a single particle measurement, then one cbit from Alice to Bob  will result 
in a qubit or a qubit  up to a rotation operator at Bob's place. If Alice
finds $\ver \Psi \ra_1$, then sending of one cbit will yield an exact a
complement qubit or a complement-qubit state up to a rotation operator.
The overall rotation operators that Bob has to apply to get a
qubit depends on the type of entangled state they have shared initially.
The following discussions refer to the case when Alice and Bob share an EPR
singlet. If Alice  chooses to prepare a real qubit, i.e.,
$\ver \Psi \ra = \cos \frac{\theta}{2} \ver 0 \ra + \sin \frac{\theta}{2}
\ver 1 \ra $, which means
on the projective Hilbert space $S^2$ the point lies on the polar line,
then the azimuthal angle $\phi$ is zero.  In this case Bob  just has to
perform a rotation (i.e. apply $\sigma_x \sigma_y$ ) or do nothing after
receiving the classical information from Alice. Alternatively, Alice could
chose to prepare a qubit chosen from equatorial line on Bloch sphere such as
$\ver \Psi \ra = \frac{1}{\sqrt 2} ( \ver 0 \ra + e^{i \phi} \ver 1 \ra )$ with
$\theta = \pi /2$. In this case when Bob gets $\ver \Psi_{\perp} \ra_{2}
=  \frac{1}{\sqrt 2} (\ver 0 \ra -  e^{i \phi} \ver 1 \ra )$ then he can still
get $ \ver \Psi \ra $ by applying $\sigma_z$.
Therefore, when the measurement outcome is  $\ver \Psi \ra_1$ or $\ver
\Psi_{\perp} \ra_1$ ( in the both cases) Bob's particle is prepared in
the (un)known state.
Thus for any real qubit our simple scheme remotely prepares a known
 state
with certainty. Since a real qubit requires a single infinity of
bits of information (as one real number $\theta$ or $\phi$ is necessary) to be send 
across a classical channel, use of shared entanglement reduces it to sending
just one cbit across a classical channel and this can be done with certainty.
For an arbitrary but {\em known} qubit  this protocol is able to transmit
half-of the time. This is because Bob cannot convert the
orthogonal-complement qubit (which he gets half of the time) since
it is {\em unknown} to him. We  know that an arbitrary unknown state cannot be
complemented \cite{akp,bhw,gp}. Though we can design a NOT gate which
 can take
$\ver 0 \ra \rightarrow \ver 1 \ra$ and $\ver 1 \ra \rightarrow \ver 0 \ra$
there is no universal NOT gate which can take $\ver \Psi \ra \rightarrow \ver
\Psi_{\perp} \ra$ as it involves an {\em anti-unitary} operation. Thus,
a doubly infinity of bits of information cannot be passed all the time with
the use of entanglement by sending just one cbit.

This shows that to remotely prepare a {\em known} qubit chosen from special
ensemble one need not do a
Bell-state measurement and send two cbits. Only single particle measurement and
one cbit is necessary from Alice to Bob, provide they share
an entangled state. In ``classical teleportation'' of a qubit it is aimed
to simulate any possible measurement on the qubit sent to Bob
(unknown to him) \cite{cgm}. One may tend to think
that since in our scheme we can remotely prepare an arbitrary {\em known}
state one half of the time Bob might not be able to simulate the measurement
statistics all the time ( as Bob cannot get a {\em unknown} qubit from the
complement qubit). However,
there is no problem with Bob for simulating the measurement statistics  on
the complement qubit (also called time-reversed qubit). This is because
the quantum mechanical probabilities and transition probabilities are
invariant under unitary and anti-unitary operations
(thanks to Wigner's theorem). This says that for any two non-orthogonal rays
$\ver \la \Psi \ver \Phi \ra \ver^2
=  \ver \la \Psi' \ver \Phi' \ra \ver^2$ where $\ver \Psi' \ra, \ver \Phi' \ra$
are related to $\ver \Psi \ra, \ver \Phi \ra$ either by unitary or anti-unitary
transformations. For example, if Bob wants to measure an observable $({\bf b}.\sigma)$,
then the probability of measurement outcome in the state $\rho = \ver \Psi \ra
\la \Psi \ver = \frac{1}{2} (1 + {\bf n}. \sigma) $ is given by

\begin{eqnarray}
P_{\pm}( \rho ) = {\rm tr}(P_{\pm}({\bf b}) \rho ) =
\frac{1}{2}(1 \pm {\bf b}. {\bf n} ),
\end{eqnarray}
where the projection operator 
$P_{\pm}({\bf b}) = \frac{1}{2}(1 \pm {\bf b}. \sigma )$. But suppose Bob gets
$\rho_{\perp} = \ver \Psi_{\perp} \ra \la \Psi_{\perp} \ver =
\frac{1}{2} (1 - {\bf n}. \sigma) $. In this case
the measurement gives a result

\begin{eqnarray}
P_{\pm}( \rho_{\perp} ) = {\rm tr}(P_{\pm}({\bf b}) \rho_{\perp} ) =
\frac{1}{2}(1 \mp {\bf b}. {\bf n} )
\end{eqnarray}
whch is different than (11).
However, Bob can always chose his apparatus such that he can make
$P_{\pm}( \rho )= P_{\pm}( \rho_{\perp} )$.
So even if Bob cannot get a qubit from a complement qubit (half of the time)
still he can get the
same measurement outcomes from it. Therefore, Bob can simulate with $100\%$
efficiency the statistics of his measurements on a qubit {\em known} to Alice
but {\em unknown} to him, provided they share an EPR pair and communicate
one cbit. Whether one can always remotely simulate all the measurement
results for a higher dimensional quantum systems is still an open
question.

Our observation also shows that the extra cbits required in  a hidden variable scenario is
1.19 and not just .19 bits as mentioned in \cite{cgm}.
So to fill the gap between LHV and quantum theory 1.19 cbits are necessary
(for lower dimensional Hilbert spaces). It should be remarked that the
2.19 cbit needed in classical teleportation protocol \cite{cgm} is not
optimal. If a better protocol exists then that will bring down the cbit
cost. We can formally state our last result in the following theorem.

{\em Theorem}:
Any LHV model which simulates teleportation of a {\em known} qubit without
entanglement will require at least one cbit (because no LHV can beat the
use of entanglement) to be transmitted from Alice to Bob.

Entanglement channel is a {\em passive} communication channel which on its
own cannot be used  for communication purposes. Supplemented with cbits it
become {\em active}, so we can regard cbits as the {\em ``soul'' of entanglement channel}.
Thus we can say that the minimum cbits required
to remotely prepare a real {\em known} qubit is one cbit (using shared entanglement)
where as to transmit an {\em unknown}
qubit one needs two cbits (as in teleportation protocol). The scenario presented
here is also very useful in the context of ``assisted cloning'' 
and ``orthogonal complementing'' of unknown states \cite{akp1}. The present
result is important, because it sets a (lower) bound on the number of cbits
required to send a {\em known} qubit using LHV. Recently, classical communication cost of remote state preparation
and distributed quantum information has been studied by Lo \cite{hklo}.
In an important paper Bennett {\it et al} \cite{bdstw} have shown that
assymptotically one needs one cbit per qubit for remote state preparation
of any qubit and have studied the remote preparation of entangled states.

\vskip .5cm

 I thank N. J. Cerf for useful discussions.
 I thank H. K. Lo for his comments and suggesting a suitable title.
 I also thank B. Terhal and D. DiVincenzo for useful remarks.
 I gratefully acknowledge the financial support from European Physical Science
 Research council (EPSRC).

\renewcommand{\baselinestretch}{1}

\end{multicols}

\newpage


\noindent
\begin{thebibliography}{99}

\bibitem{cb} C. Bennett, G. Brassard, C. Crepeau, R. Jozsa, A. Peres,
 and W. K. Wooters, Phys. Rev. Lett. {\bf 70}, 1895 (1993).

\bibitem{db} D. Bouwmeester, J. W. Pan, K. Mattle, M. Eibl, H. Weinfurter,
and A. Zeilinger, Nature {\bf 390}, 575 (1997).

\bibitem{dbe} D. Boschi, S. Branca, F. De Martini, L. Hardy and
S. Popescu, Phys. Rev. Lett. {\bf 80}, 1121 (1998).

\bibitem{af} A. Furusawa, J. L. Sorensen, S. L. Braunstein, C. A. Fuchs,
H. J. Kimble and  E. S. Polzik, Science {\bf 282}, 706 (1998).


\bibitem{rj} R. Jozsa, {\sl Geometric Issues in Foundations of Science},
Eds. S. Huggett et al, Oxford Univ. Press, 1997.

\bibitem{lv} L. Vaidman, LANL report, quant-ph/9810089.

\bibitem{sp} S. Popescu, Phys. Rev. Lett. {\bf 74}, 2619 (1995).

\bibitem{lh} L. Hardy, LANL report, quant-ph/9906123.

\bibitem{betal} G. Brassard, R. Cleve and A. Tapp, LANL report, quant-ph/9901035.

\bibitem{ms} M. Steiner, LANL report, quant-ph/9902014.

\bibitem{cgm} N. J. Cerf, N. Gisin, and S. Massar, LANL report, quant-ph/9906105.


\bibitem{akp} A. K. Pati, {\it  New limitations in quantum information processing},
  Preprint (1999).

\bibitem{bhw} V. Bu\v{z}ek, M. Hillery and R. F. Werner, LANL Report,
 quant-ph/9901053.

\bibitem{gp} N. Gisin and S. Popescu, LANL Report, quant-ph/9901072.

\bibitem{akp1} A. K. Pati, Phys. Rev. A, {\bf 61}, 022308 (2000).

\bibitem{hklo} H. K. Lo, LANL report, quant-ph/9912009.
\bibitem{bdstw} C. H. Bennett, D. P. DiVincenzo, J. A. Smolin, B. M. Terhal,
and W. K. Wootters, LANL report, quant-ph/0006044.

\end{thebibliography}
\end{document}